\documentclass[
superscriptaddress,
preprint,
amsmath,amssymb,
aps,
floatfix,
]{revtex4-1}

\usepackage{graphicx}
\usepackage{dcolumn}
\usepackage{bm}
\usepackage[mathlines]{lineno}

\begin{document}

\title{Interaction of Airy-Gaussian beams in defected photonic lattices}
\author{Zhiwei Shi}
\thanks{Corresponding author: szwstar@gdut.edu.cn}
\affiliation{School of Electro-mechanical Engineering, Guangdong University of Technology, Guangzhou 510006,P.R.China}
\author{Jing Xue}
\affiliation{School of Information Engineering, Guangdong University of Technology, Guangzhou 510006,P.R.China}
\author{Xing Zhu}
\affiliation{Department of Physics, Guangdong University of Education , Guangzhou 510303, China}
\author{Ying Xiang}
\affiliation{School of Information Engineering, Guangdong University of Technology, Guangzhou 510006,P.R.China}
\author{Huagang Li}
\thanks{Corresponding author: lihuagang@gdei.edu.cn}
\affiliation{Department of Physics, Guangdong University of Education , Guangzhou 510303, China}
\begin{abstract}
We investigate interactions by means of direct numerical simulations between two finite Airy-Gaussian (AiG) beams in different media with the defected photonic lattices in one transverse dimension. We discuss different lattice structures in which the beams with different intensities and phases are launched into the medium, but accelerate in opposite directions. During interactions we see the interference fringe, breathers and soliton pairs generated that are not accelerating. In the linear media, the initial deflection direction of the accelerated beams is changed by adjusting the phase shift and the beam interval. For a certain lattice period, the periodic interference fringe can form. A constructive or destructive interference can vary with the defect depth and phase shift. While the nonlinearity is introduced, the breathers is generated. Especially, the appropriate beam amplitude and lattice depth may lead to the formation of soliton pairs.
\end{abstract}

\maketitle
\section{Introduction}

Photonics and especially nonlinear photonics has experienced rapid development over the last few decades. In this growth, the novel accelerated Airy beams played one important role and nonlinear effects such as soliton formation was the other key point. New fields of research has been opened completely because of combining both topics to study nonlinear interaction of accelerated beams. The Airy quantum wave packet of infinite extent was introduced by Berry and Balazs as free particle solutions of the Schr\"{o}dinger equation~\cite{1}. However, the initial Airy function is not realizable in practice.

In optics, because they can be realized in practice and exhibit self-accelerating,
non-diffracting, and self-healing properties during propagation, the Airy optical beams which retain a finite energy were widely investigated theoretically and experimentally since Siviloglou \textsl{et al}~\cite{2,3} discussed them by using different methods for the first time in 2007. In the past decade, such self-accelerating optical beams have been studied, mostly in uniform media which include linear media~\cite{2,3,4,5,6,7,8,9,10}, Kerr nonlinear dielectrics~\cite{11,12,13,14,15}, photorefractive media~\cite{17}, nonlocal nonlinear media~\cite{18,19}, and quadratic media~\cite{11,22}. Because of the existence of nonlinearity, optical solitons can be formed with the Airy beams in different nonlinear media~\cite{24,25,26,27,28}. At the same time, a few people studied the propagation properties of optical Airy beams in the nonuniform structures~\cite{29,30,31}. In 2005, Christodoulides et al. identified nondiffracting beams in two-dimensional periodic systems, exhibiting symmetry properties and phase structure characteristic of the band(s) they are associated with~\cite{29}. Z. Chen \textsl{et al} studied the behavior of Airy beams propagating from a nonlinear medium to a linear mediumn in 2010~\cite{30} and demonstrated both experimentally and theoretically controlled acceleration of one- and two-dimensional Airy
beams in optically induced refractive-index potentials in 2010~\cite{31}, respectively.
Recently, Dragana M Jovi\'{c} \textsl{et al} reported that the propagation dynamics and
beam acceleration are controlled with positive and negative defects, and appropriate refractive index change~\cite{32}. Moreover, they analyzed how an optically induced photonic lattice affects and modifies the acceleration of Airy beams~\cite{33} and demonstrated control over the acceleration of two-dimensional Airy beams propagating in optically induced
photonic lattices~\cite{34}.

As a generalized form of the Airy beams, AiG beams can carry finite energy and maintain the diffraction-free propagation properties within a finite propagation distance~\cite{35}. Many researchers have studied the AiG beams both theoretically and experimentally~\cite{35,36,37,38,40,41}. Deng \textsl{et al} have theoretically investigated the propagation of the AiG beam in uniaxial crystals~\cite{36}, strongly nonlocal nonlinear media~\cite{37}, Kerr media~\cite{38}. Ez-Zariy~\cite{40} and Zhou~\cite{41} \textsl{et al} have discussed propagation characteristics of finite Airy-Gaussian beams through an apertured misaligned first order ABCD optical system and the fractional Fourier transform plane, respectively. Similarly asymptotic preservation of a self-accelerating property is observed with AiG beams in nonlinear media and with Airy beam introduced in defected waveguide arrays.

Above-mentioned papers have investigated dynamics and properties of single accelerating beams. Moreover, similar to the interactions of solitons, interactions between Airy beams have gradually attracted attention of researchers. Interactions between Airy
pulse and temporal solitons at the same
center wavelength~\cite{42} or at a different wavelength~\cite{43} were studied.
The interaction of an accelerating Airy beam and a solitary wave was also investigated in various media~\cite{44}. Wolfersberger \textsl{et al} analysed the dynamics of two incoherent counterpropagating Airy beams interacting in a photorefractive crystal under focusing conditions~\cite{45,46}. Zhang \textsl{et al}~\cite{47,48} and Deng \textsl{et al}~\cite{49,50} studied numerically interactions of Airy and AiG beams in nonlinear media in one transverse dimension, respectively. Based on the effect of nonlocality, Shen \textsl{et al} obtained stationary bound states of in-phase as well as out-of-phase Airy beams in nonlocal nonlinear media~\cite{51,52}. The interaction between a broad accelerating Airy beam and an intense Gaussian beam was also investigated numerically and experimentally to demonstrate gravitational dynamics in a nonlocal thermal nonlinearity~\cite{53}. Thus far, interactions of self-accelerating beams in uniform media were reported. However, interactions of self-accelerating beams in the waveguide arrays, especially in the defected photonic lattices,  have not been mentioned.

In this paper, we will numerically study the interactions dynamics of AiG beams in one-dimensional photonic lattices including defects. We realize different defect lattices by embedding the positive and negative defects into the regular lattice and research the influence of the different physical parameters on the AiG beam interaction. The organization of the paper is as follows. We briefly introduce the theoretical model and basic equations in
Sec.~\ref{section:two}; in Sec.~\ref{section:three}, we discuss numerically interactions of two AiG beams in defected photonic lattices in details. Section~\ref{section:four} concludes the paper.
\begin{figure}[htb]
\centering
\fbox{\includegraphics[width=\linewidth]{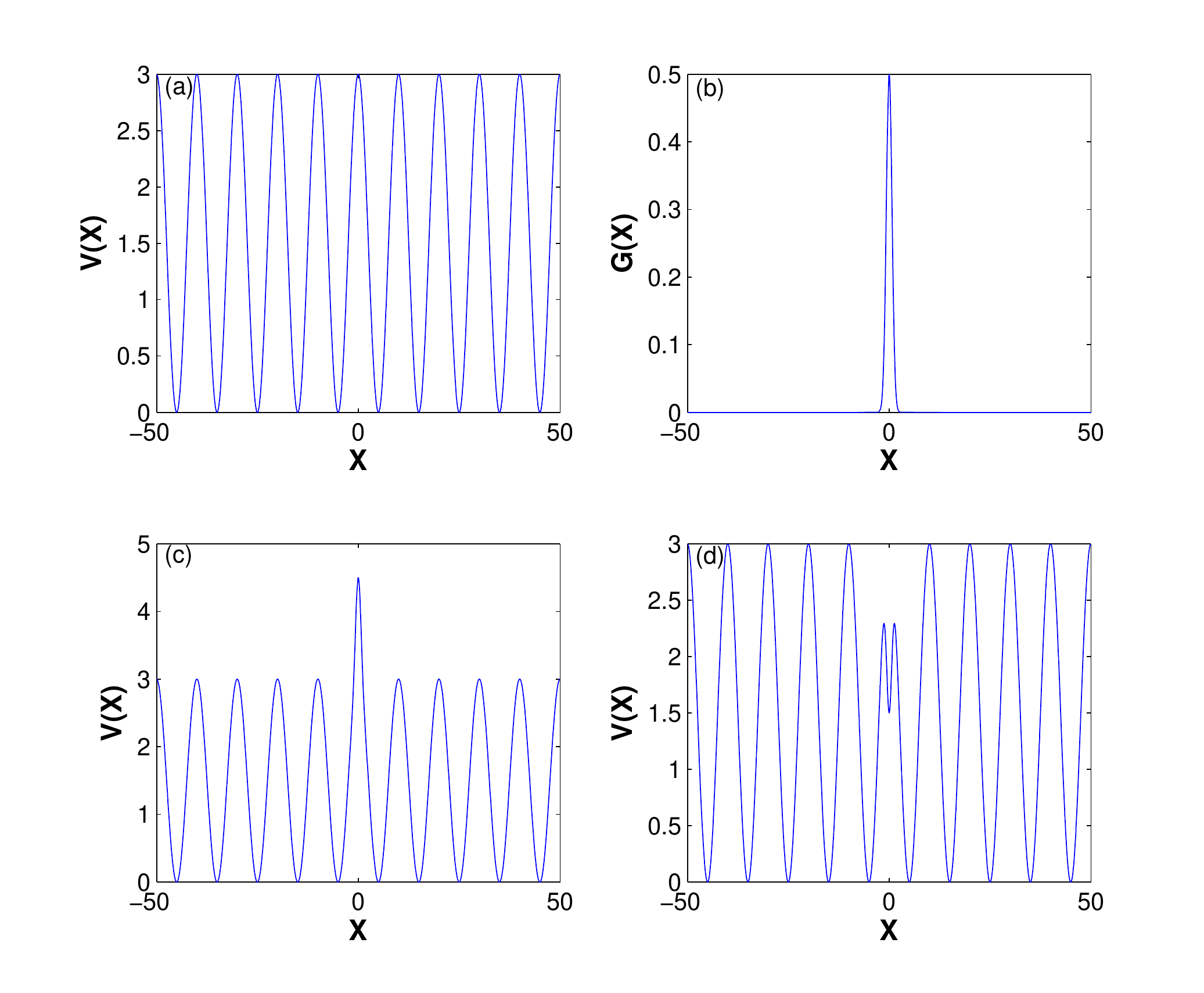}}
\caption{(Color online) Defect generation in optical photonic lattice. (a) The regular lattice distribution $(\delta n=0)$. (b) The Gaussian beam intensity distribution $G(X)$. Numerical realization of (c) positive ($\delta n=0.5$) and (d) negative ($\delta n=-0.5$) defect lattices. The other physical parameters $A_1=A_2=2$, $A_n=3$, $T=0.1$mm, $\delta\phi=0$, $D=3$, $Q=0.05$, and $\gamma=0$.}
\label{fig:one}
\end{figure}
\section{The theoretical model and basic equations}
\label{section:two}
To study the interaction characteristics of AiG beams in defected photonic lattices, along the propagation distance $z$, we consider that the scale equation for the propagation of a slowly varyinng envelope $q$ of the optical electric field in one transverse dimension in the paraxial approximation is of the nonlinear Schr\"{o}dinger equation
\begin{equation}
i\frac{\partial q(X,Z)}{\partial Z}+\frac{1}{2}\frac{\partial^2q(X,Z)}{\partial X^2}+V(X)q(X,Z)+\gamma|q(X,Z)|^2q(X,Z)=0,
\label{eq:one}
\end{equation}
\begin{figure}[htb]
\centering
\fbox{\includegraphics[width=\linewidth]{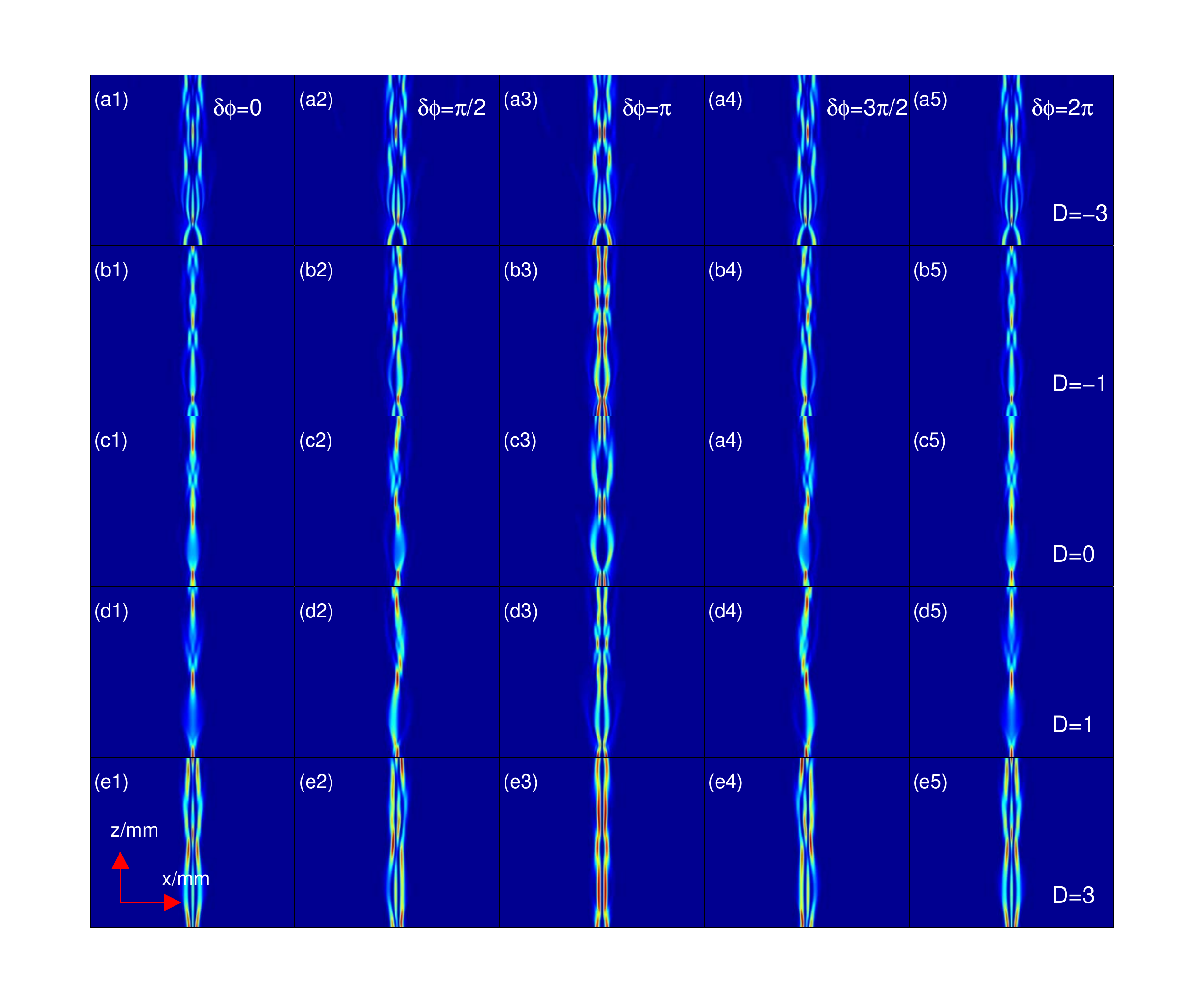}}
\caption{(color online) The interaction of two AiG beams with changing the values of $\delta\phi$ and $D$ for $A_1=A_2=2$, $A_n=3$, $T=0.2$mm, $\delta n=0$, $Q=0.05$, and $\gamma=0$.}
\label{fig:two}
\end{figure}
where $X=x/w_0$ is the dimensionless transverse coordinates scaled by the characteristic length $w_0$, $Z=z/kw^2_0$ with $k=2\pi/\lambda$, $V(X)=A_ncos^2(\pi Xw_0/T)(1+\delta n\exp(-X^2))$ is the periodic refractive-index profile of the array with the lattice period $T$, $A_n$ is the lattice modulation depth, and $\delta n$ is the defect depth. Here, we assume $w_0=10\mu$m and the wave length $\lambda=600$nm. In Kerr media, the beams are self-focusing ($\gamma=1$) when the nonlinear refractive index is greater than zero and self-defocusing ($\gamma=-1$) when the nonlinear refractive index is smaller than zero. It is well known that the spatial solitons can be steadily transmitting in the (1+1)D local Kerr medium when the nonlinear effect balances the diffraction effect. Considering a AiG beam, its initial field distribution can be read as~\cite{36,37,38,40,41,49,50}
\begin{equation}
q(X,0)=A_0Ai(X)\exp(\alpha X)\exp(-QX^2),
\label{eq:two}
\end{equation}
\begin{figure}[htb]
\centering
\fbox{\includegraphics[width=\linewidth]{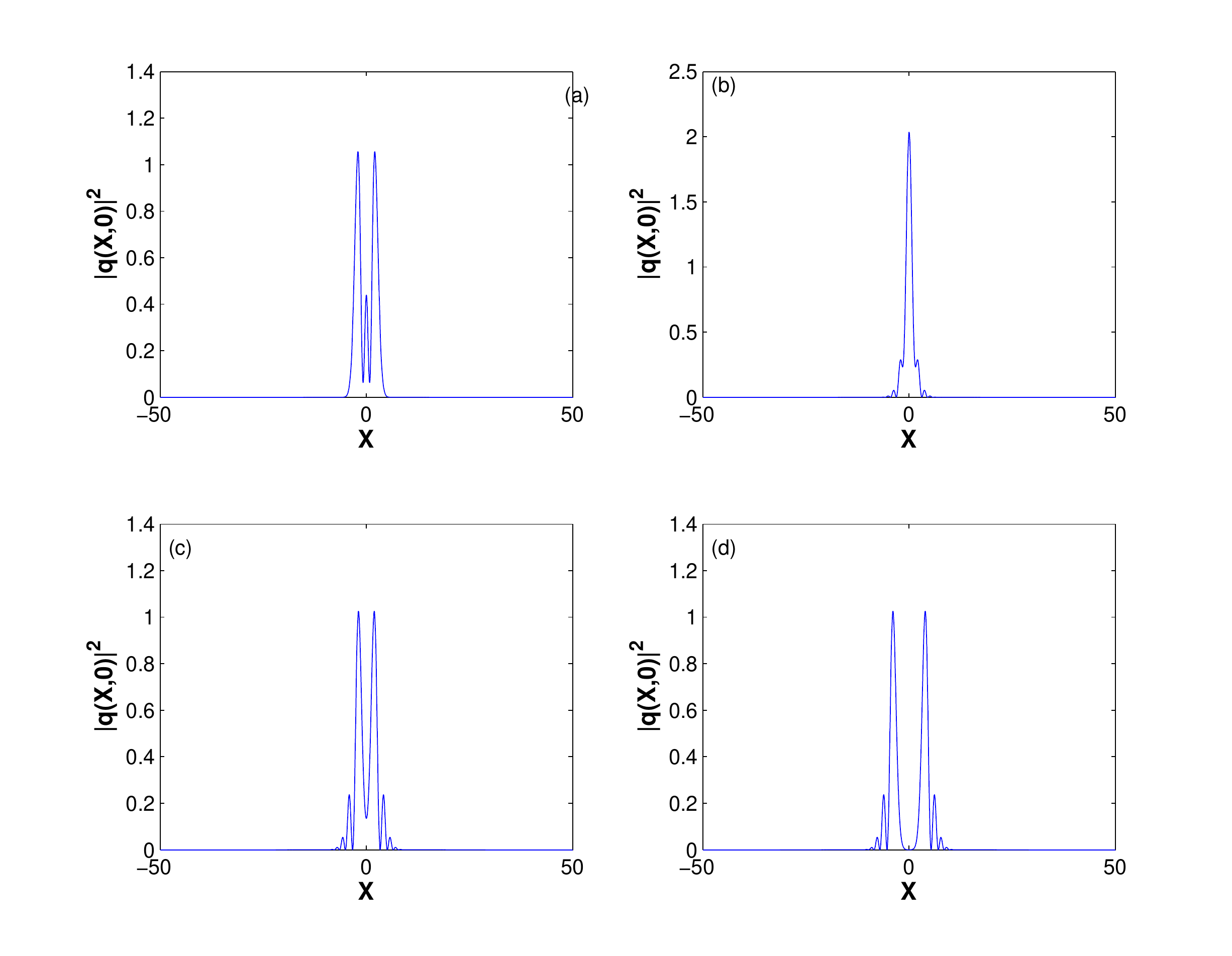}}
\caption{(color online) Intensity profiles of AiG beams with $D=3$ (a), $D=1$ (b), $D=-1$ (c), and $D=-3$ (d) for $\delta\phi=\pi/2$. The other physical parameters are the same as Fig.~\ref{fig:two}.}
\label{fig:three}
\end{figure}
where $A_0$ denotes the constant amplitude, $Ai(\cdot)$ is the Airy function, $\alpha=0.01$ in the exponential function is a parameter associated with the truncation of the AiG beams, and $Q$ is the distribution factor controlling
the beam that will tend to the Gaussian beam with a larger value and the Airy beam with a smaller value. The expression~(\ref{eq:two}) of the initial field $q$ is the single-beam solutions of Eq.~(\ref{eq:one}). To investigate the AiG beam interactions, we should construct more complex incident beams, made up of two shifted single beams, launched in parallel but accelerating in opposite directions. Thus, we assume that the incident beam will be composed of two shifted AiG beams with a fixed relative phase and different amplitudes between them~\cite{47,48,49,50,51,52},
\begin{eqnarray}
q(X,0)=A_1Ai(X-D)\exp(\alpha (X-D))\exp(-Q(X-D)^2)\nonumber\\
+A_2Ai(-X-D)\exp(\alpha (-X-D))\exp(-Q(-X-D)^2)\exp(i\delta\phi).
\label{eq:three}
\end{eqnarray}
where $A_1$ and $A_2$ are the amplitudes of the two AiG beams, $D$ is the parameter controlling the beam separations, and $\delta\phi$ is the parameter
controlling the phase shift with $\delta\phi=0$ and $\delta\phi=\pi$ describing in-phase and out-of-phase AiG beams, respectively. To investigate the interaction of the two AiG beams for different beam factors, we have implemented comprehensive split-step Fourier methods to solve Eq.~(\ref{eq:one}) and model the light propagation in defected photonic lattices $V$. The propagation equation~(\ref{eq:one}) is evaluated numerically, taking Eq.~(\ref{eq:three}) as the initial input AiG beam. Figure \ref{fig:one} shows the basic scheme of the defect realization using a Gaussian beam. The regular lattice distribution ($\delta n=0$) and Gaussian beam intensity distribution $G(X)=\delta n\exp(-(X^2))$ are illustrated in Figs.~\ref{fig:one}(a) and ~\ref{fig:one}(b), respectively. Figs.~\ref{fig:one}(c) and~\ref{fig:one}(d) show the calculated refractive index modulation results for both the positive ($\delta n=0.5$) and negative ($\delta n=-0.5$) defect lattices.

\section{The numerical results of interacting AiG beams in defected photonic lattices}
\label{section:three}

\begin{figure}[htb]
\centering
\fbox{\includegraphics[height=10cm,width=\linewidth]{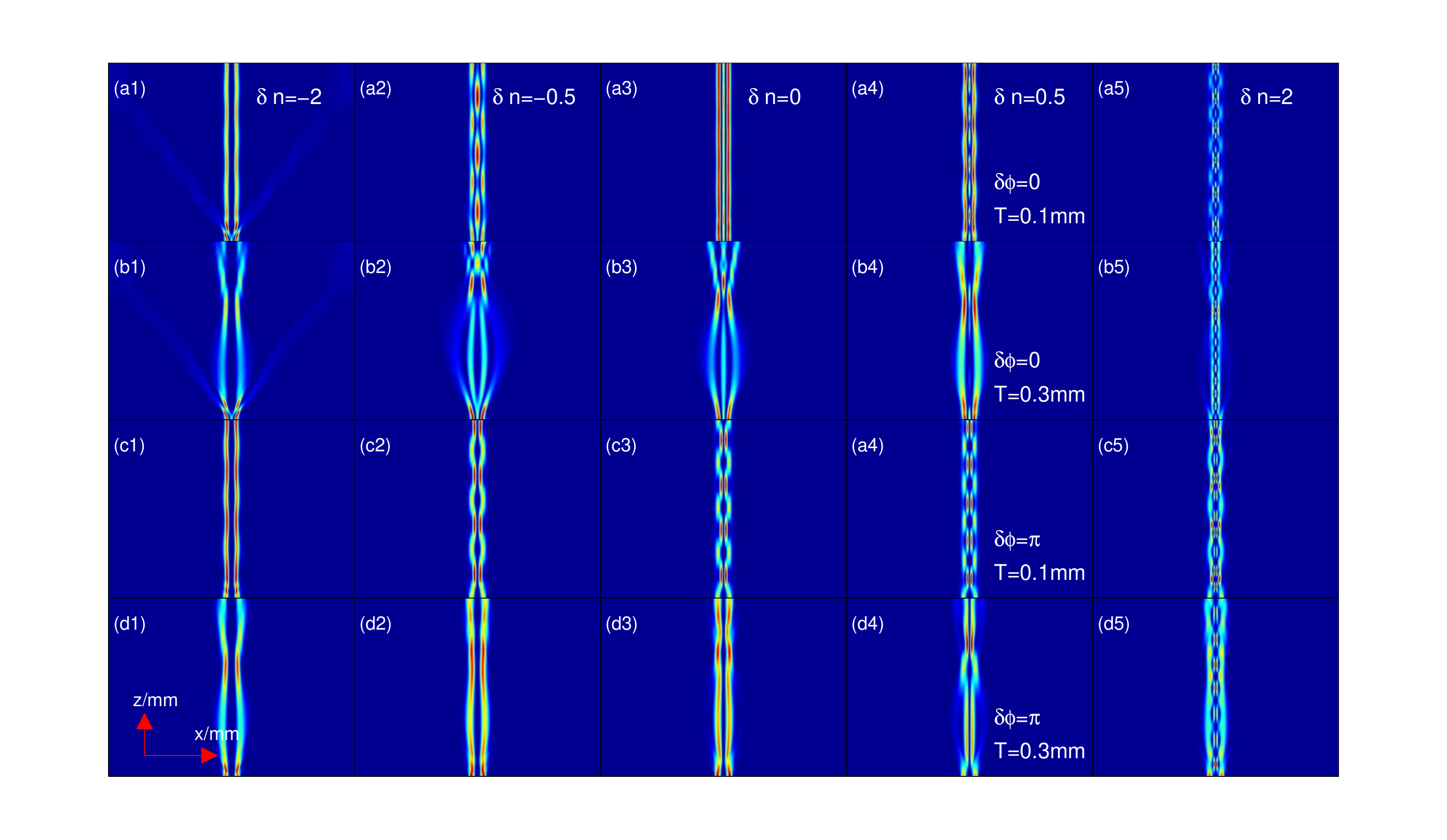}}
\caption{(color online) The interaction of two AiG beams with changing the values of $\delta\phi$, $\delta n$, and $T$ for $A_1=A_2=2$, $A_n=3$, $D=3$, $Q=0.05$, and $\gamma=0$.}
\label{fig:four}
\end{figure}
If we assume $\gamma=0$ and $\delta n=0$ in Eq.~(\ref{eq:one}), the
``interaction" is actually a linear interference in the regular photonic lattices. We display the evolution of the incidence from Eq.~(\ref{eq:three}), for different $D$ and $\delta\phi$ in Fig.~\ref{fig:two}. The results shown in Figs.~\ref{fig:two}(a1)-\ref{fig:two}(e1), \ref{fig:two}(a3)-\ref{fig:two}(e3), and \ref{fig:two}(a5)-\ref{fig:two}(e5) are completely different from Fig.~\ref{fig:two} in Ref.~\cite{48} because of the exsitence of the photonic lattices. However, the behavior of the central interference is similar. The central interference fringe in the in-phase case (Figs.~\ref{fig:two}(a1)-\ref{fig:two}(e1) and \ref{fig:two}(a5)-\ref{fig:two}(e5)) is bright, whereas in the out-of-phase case (\ref{fig:two}(a3)-\ref{fig:two}(e3)) it is dark, as it should be for
a constructive and destructive interference~\cite{48}. Some pseudo-periodic mutual-focusing can be observed in the central region, especially when the interval of beams is closer, such as $D=0$ (see \ref{fig:two}(c1)-\ref{fig:two}(c5)). This results from the diffraction, superposition, and interference of the curved accelerating beams as the beams propagate in the photonic lattices. Of course, no breathers can form.
\begin{figure}[htb]
\centering
\fbox{\includegraphics[width=\linewidth]{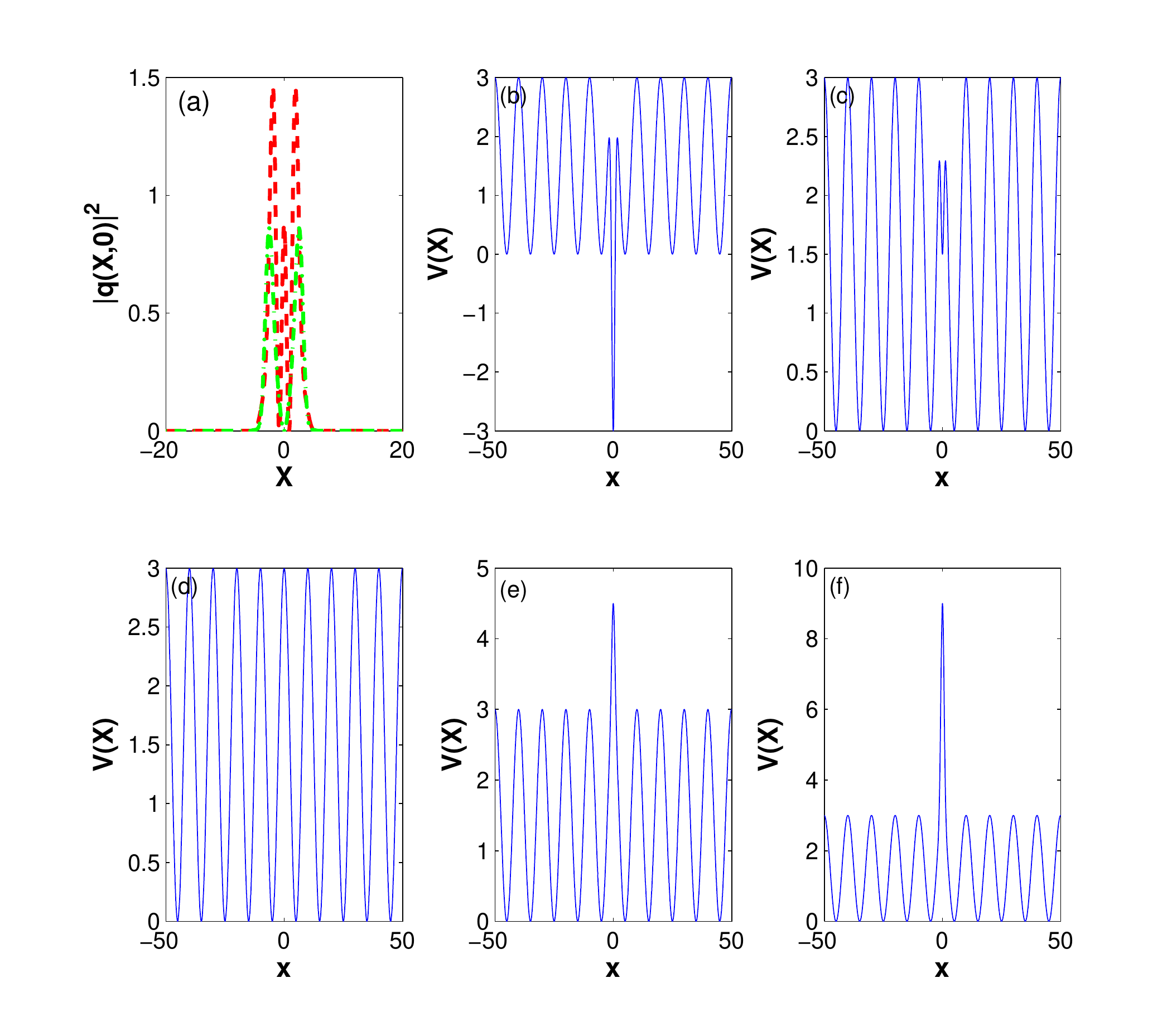}}
\caption{(color online) (a) Intensity profiles of the initial input AiG beams with $\delta\phi=0$ (red dashed line) and $\delta\phi=\pi$ (green dash-dotted line) for $\delta n=-2$ and $T=0.1$mm. The defected photonic lattices with $\delta n=-2$ (b), $\delta n=-0.5$ (c), $\delta n=0$ (d), $\delta n=0.5$ (e), and $\delta n=2$ (f) for $\delta\phi=0$ or $\delta\phi=\pi$ and $T=0.1$mm. The other physical parameters are the same as Fig.~\ref{fig:four}.}
\label{fig:five}
\end{figure}
\begin{figure}[htb]
\centering
\fbox{\includegraphics[width=\linewidth]{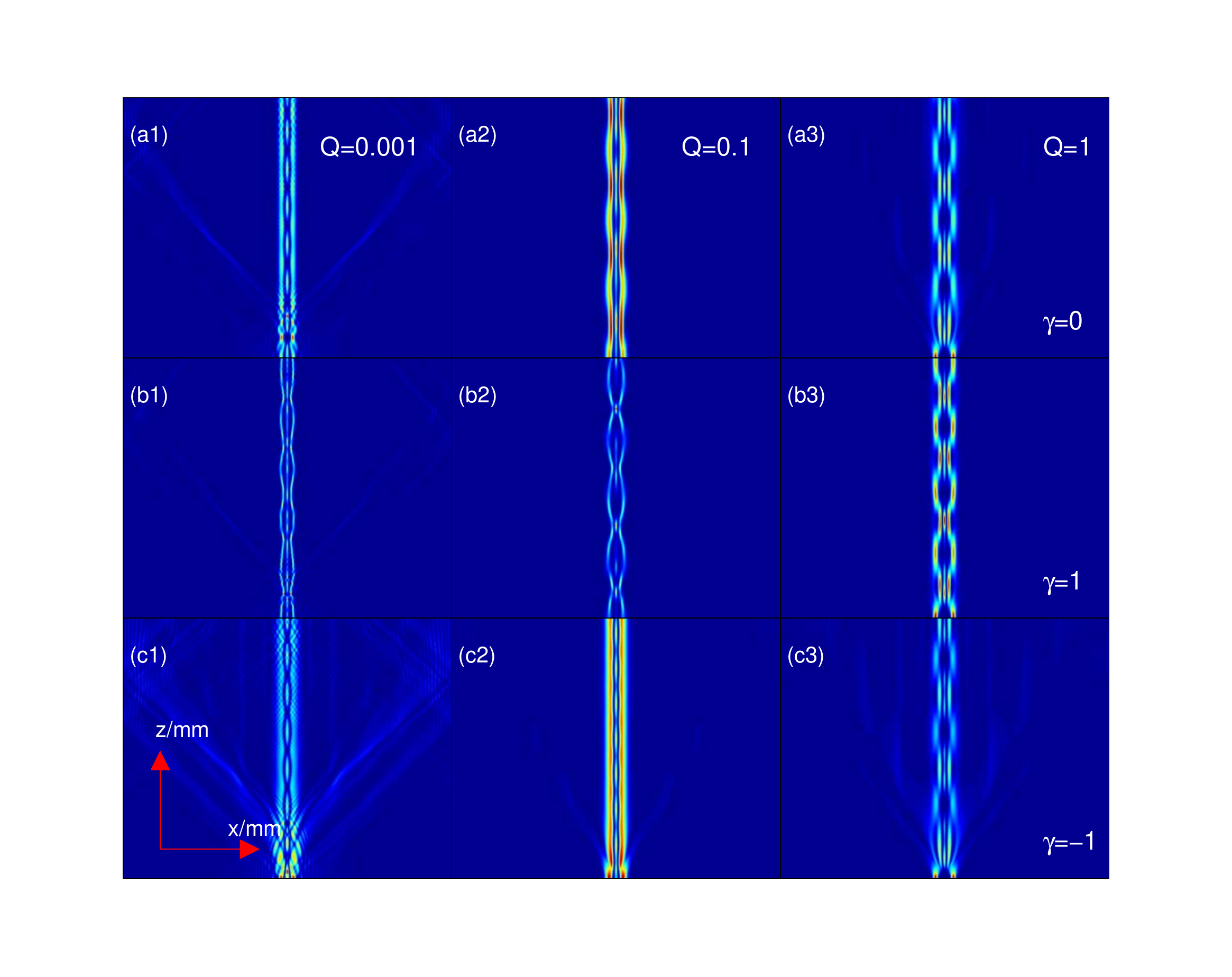}}
\caption{(color online) The interaction of two AiG beams with changing the values of $Q$, and $\gamma$ for $A_1=A_2=4$, $A_n=3$, $D=3$, $Q=0.05$, $\delta\phi=0$, $T=0.1$mm, and $\delta n=0.5$.}
\label{fig:six}
\end{figure}

When we change the phase shift $\delta\phi$, the beam direction varies simultaneously because the change of $\delta\phi$ influences the interaction of two beams. If $\delta\phi=\pi/2$, the direction of the accelerated beams will be firstly turned to the right; If $\delta\phi=3\pi/2$, the accelerated beams will firstly go to the opposite direction. However, it is noted that the case is inverse when $D=1$. This can be interpreted by Fig.~\ref{fig:three}. We can see that the intensity profile and value of the AiG beam with $D=1$ are different from the other three profiles and values. Thus, the initial deflection direction of the accelerated beam with $D=1$ is firstly left based on the interaction between the beam and lattice.
As a result, we can say that the propagation direction of the beams can be changed by adjusting the values of the phase shift $\delta\phi$ and the beam interval $D$. Interestingly, the interaction of the two beams is the strongest with both in-phase and out-of-phase situations, so we will next study the interaction of the beams about two key cases.

\begin{figure}[htb]
\centering
\fbox{\includegraphics[width=\linewidth]{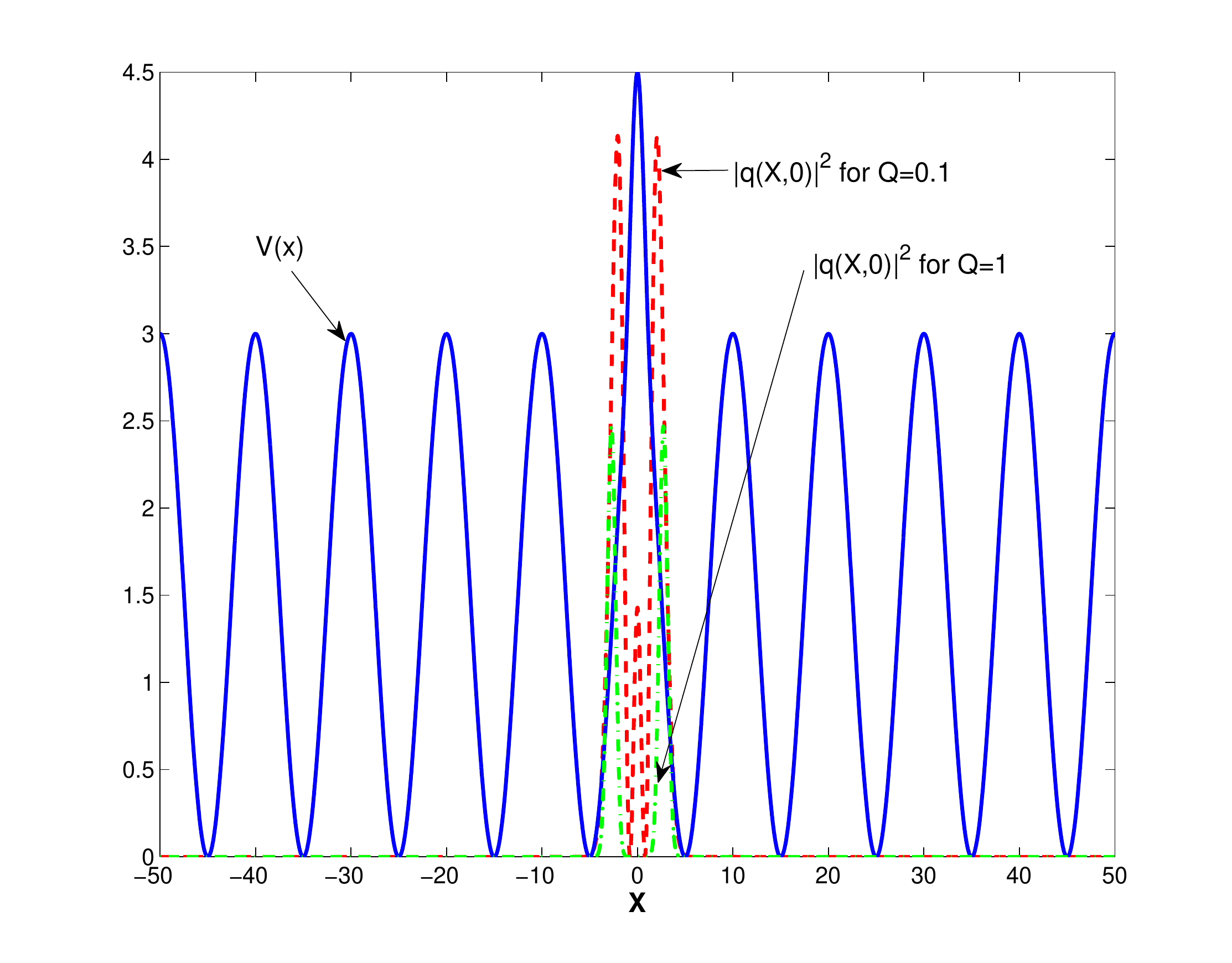}}
\caption{(color online) Intensity profile of AiG beams $|q(X,0)|^2$ and the defected lattice $V(X)$. The red dashed and green dash-dotted lines denote $|q(X,0)|^2$ at $Q=0.1$ and $Q=1$, respectively. The blue solid line shows $V(X)$. The other physical parameters are the same as Fig.~\ref{fig:six}.}
\label{fig:seven}
\end{figure}
\begin{figure}[htb]
\centering
\fbox{\includegraphics[width=\linewidth]{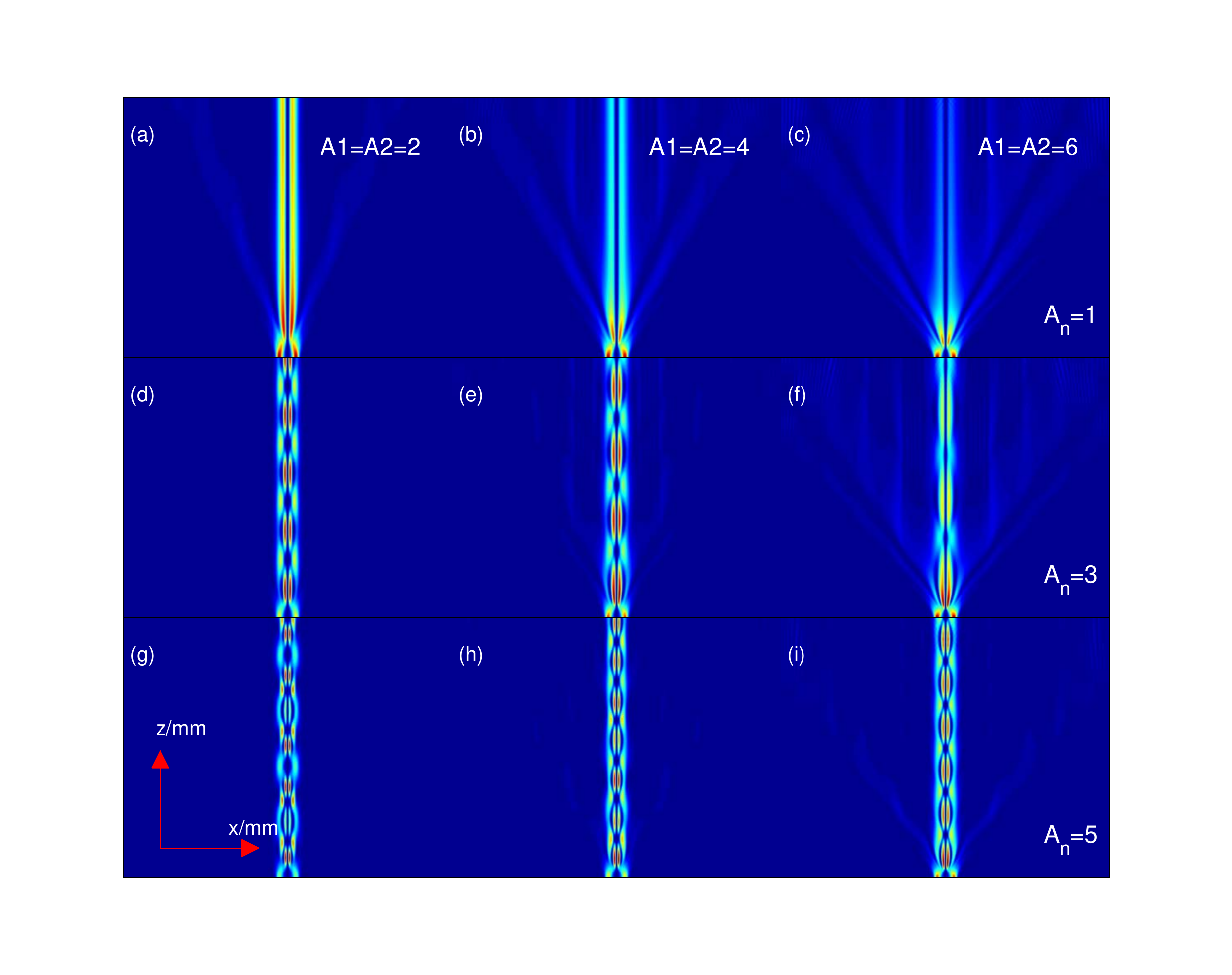}}
\caption{(color online) The interaction of two AiG beams with changing the values of $A_1$, $A_2$, and $A_n$ for $T=0.1$mm, $\gamma=-1$, $D=3$, $Q=0.05$, $\delta\phi=\pi$, and $\delta n=0.5$.}
\label{fig:eight}
\end{figure}
\begin{figure}[htb]
\centering
\fbox{\includegraphics[width=\linewidth]{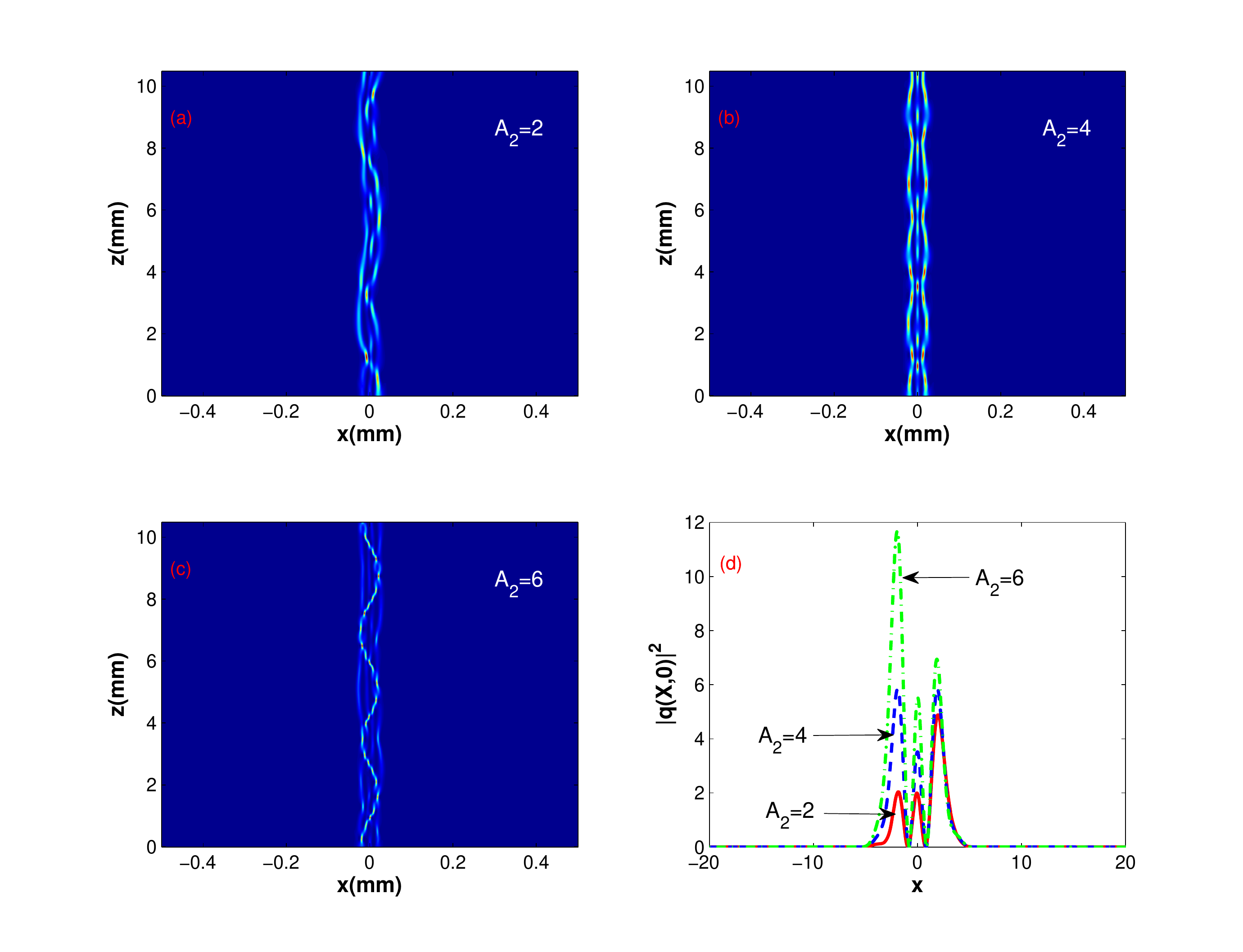}}
\caption{(color online) The interaction (a)-(c) and intensity profiles of two AiG beams with changing the values of $A_2$ for $A_1=4$, $A_n=3$, $T=0.1$mm, $\gamma=1$, $D=3$, $Q=0.05$, $\delta\phi=0$, and $\delta n=0.5$.}
\label{fig:nine}
\end{figure}
Fig.~\ref{fig:four} shows the propagation distribution of the in-phase and out-of-phase AiG beams with the same amplitude for different $\delta\phi$, $\delta n$, and $T$. At first, we discuss that the influence of the defect depth $\delta n$ on the interaction of AiG beams (red dashed line in Fig.~\ref{fig:five}(a)) when the other parameters are constant. At $\delta\phi=0$ and $T=0.1$mm, as shown in Figs.~\ref{fig:four}(a1)-\ref{fig:four}(a5), the interaction of the in-phase beams varies with $\delta n$. One can see that the ``interaction" forms a linear interference in the defected photonic lattices and the central interference fringe is bright except $\delta n=-2$. It is caused by the influence of the defected lattice (see Fig.~\ref{fig:five}(b)) different from the other cases (see Figs.~\ref{fig:five}(c)-\ref{fig:five}(f)). The negative defected lattice exists negative value, so the central interference fringe is dark. The central dark interference fringe also appears in Figs.~\ref{fig:four}(c1)-\ref{fig:four}(c5), where $\delta\phi=\pi$ and $T=0.1$mm, that is to say, two beams is out-of-phase. The difference of the initial input in-phase and out-of-phase beams shows in Fig.~\ref{fig:five}(a). Especially, when $T=0.1$mm, one can find that some periodic mutual-focusing interference fringes, which are similar to breathers, take shape from Figs.~\ref{fig:four}(a1)-\ref{fig:four}(a5) and Figs.~\ref{fig:four}(c1)-\ref{fig:four}(c5). However, when $T$ is bigger, such as $T=0.3$mm shown in Figs.~\ref{fig:four}(b1)-\ref{fig:four}(b5) and Figs.~\ref{fig:four}(d1)-\ref{fig:four}(d5), the mutual-focusing interference fringes is no longer periodic. These results account for the big effect of the defected lattices on the beam interaction.

Fig.~\ref{fig:six} shows the interaction of two in-phase AiG beams with
the same amplitude for the different distribution factor $Q$ and nonlinear parameter $\gamma$.
The mutual-focusing distributions in the central regions are periodic for all cases, if we don't consider the influence of the numerical integration window on the distribution for $Q=0.001$ in Figs.~\ref{fig:six}(a1)-\ref{fig:six}(c1).
In a linear medium ($\gamma=0$), when $Q=0.001$, the AiG beam goes to the field distribution of Airy beams which hold some side lobes, so that the defected lattice doesn't completely restrain the part energy (see Fig.~\ref{fig:six}(a1)). On the contrary, the field distribution of Gaussian beams has also been exhibited as $Q=1$ (see Fig.~\ref{fig:six}(a3)). The AiG beam is self-accelerating no longer, which is very similar to the propagation of the Gaussian beam in the linear media. It is noted that there exists some part energy in the other region apart from the central region in Fig.~\ref{fig:six}(a3) opposite to Fig.~\ref{fig:six}(a2). The reason can be explained in Fig.~\ref{fig:seven}.  At $Q=1$, we can say that two independent Gaussian beams (green dash-dotted line) interact in the defected lattice, so they also exists the part diffraction. While $Q=0.1$, the red dashed line shows that two AiG beams propagate hand-in-hand. This helps the defected lattice so that all the energy can be restricted in the central region as shown in Fig.~\ref{fig:six}(a2).
When we introduce the nonlinearity, the situation has changed. As the medium is the self-focusing nonlinear medium ($\gamma=1$), we can see that the self-focusing nonlinearity further traps some energy and breathers has formed from Figs.~\ref{fig:six}(b1)-\ref{fig:six}(b3). However, while $\gamma=-1$, the self-defocusing nonlinearity increases the diffraction. Of course, the field distributions in the central region can also be regarded as breathers as shown in Figs.~\ref{fig:six}(b1)-\ref{fig:six}(b3).

In order to further study the effect of the nonlinearity and lattice on the interaction of two AiG beams, we will change the amplitude $A_1$ and $A_2$ and the lattice depth $A_n$. Firstly, the propagation images of two out-of- phase beams show in Fig.~\ref{fig:eight} in self-defocusing media. we assume that the value of $A_1$ and $A_2$ is same and change their value simultaneously. As $A_n=1$, one can see that the soliton pairs forms at $A_1=A_2=2$ as shown in Fig.~\ref{fig:eight}(a) from the role of the repulsive force of the self-defocusing nonlinearity and constraint force of the defected lattice in the AiG beams. However, when we increase $A_1$ and $A_2$ value, in other words, the self-defocusing nonlinearity grows, the soliton pairs can no longer exist (see Figs.~\ref{fig:eight}(b) and ~\ref{fig:eight}(c)). If $A_n$ is bigger, such as $A_n=3$ or $A_n=5$, one can see that the defected lattice further constraints the beams (see Figs.~\ref{fig:eight}(d)-\ref{fig:eight}(i)). Interestingly, the breathers appears in Fig.~\ref{fig:eight}(d), Fig.~\ref{fig:eight}(e), and Figs.~\ref{fig:eight}(g)-\ref{fig:eight}(i) if we ignore the little energy dissipative energy. Secondly, we consider that the interaction of two in-phase beams for different $A_1$ and $A_2$ in self-focusing media, as shown in Fig.~\ref{fig:nine}. The asymmetrical intensity distribution of the beams is formed by the interaction of the two AiG beams with $A_1\neq A_2$. At this situation, the breathers are discovered in a symmetrical distribution which only appears on one side from Figs.~\ref{fig:nine}(a) and \ref{fig:nine}(c). In addition, one can see that the deflecting direction affected by the different intensity distributions (see Fig.~\ref{fig:nine}(d)). Compared Fig.~\ref{fig:nine}(b) with Figs.~\ref{fig:nine}(a) and \ref{fig:nine}(c), the breathers is symmetric because $A_1$ and $A_2$ are the same value.
\section{Conclusion}
\label{section:four}
To conclude, we have investigated the interactions of two different amplitude and phase AiG
beams in linear and nonlinear media with the defected lattices by using the numerical simulations with the split-step Fourier method. We find that the interference fringe, breathers and soliton pairs can be produced in these interactions.
The generated the interference fringe, breathers and soliton pairs in the central region do not accelerate transversely, because their properties are determined by the
underlying the media with defected lattices and not by the incident beam from which they are generated. In the linear media, the phase shift and the beam interval can affect the initial deflection direction of the accelerated beams. In general, the central interference fringe in the in-phase case is bright, whereas in the out-of-phase case it is dark. Interestingly, when a lattice period is appropriate, the periodic interference fringe may be formed. A constructive or destructive interference can also be influenced by the defect depth. While the nonlinearity is introduced, the breathers is generated. As the medium is the self-focusing nonlinear medium, the self-focusing nonlinearity further traps some energy and breathers has formed. Though the self-defocusing nonlinearity increases the diffraction, the field distributions in the central region can also be regarded as breathers. Especially, when we select the appropriate beam amplitude and lattice depth, soliton pairs may be shaped. In addition, the interaction of the two AiG beams with different amplitudes can lead to the asymmetrical intensity distribution of the beams.
\section*{Acknowledgments}

 This project was supported by the National Natural Science Foundation of China under (Grant No. 11547212, 11374067) and Guangdong Provincial Natural Science Foundation (Grant No. 2016A030313747), Guangdong Provincial Science and Technology Plan (Grant No. 2014A050503064, 2016A050502055).

\end{document}